# Pattern generation by dissipative parametric instability

A. M. Perego[1,2,*], N. Tarasov[1,3], D. V. Churkin[1,4,5], S. K. Turitsyn[1,4] and K. Staliunas[2,6]

[1] *Aston Institute of Photonic Technologies, Aston University, Birmingham B4 7ET, UK*
[2] *Departament de Fisica i Enginyeria Nuclear, Universitat Politècnica de Catalunya, E-08222, Barcelona, Spain*
[3] *Institute of Computational Technologies SB RAS, Novosibirsk, 630090, Russia*
[4] *Novosibirsk State University, Novosibirsk, 630090, Russia*
[5] *Institute of Automation and Electrometry SB RAS, Novosibirsk, 630090, Russia*
[6] *Institució Catalana de Recerca i Estudis Avançats, Passeig Lluis Companys 23, E-08010, Barcelona, Spain*
[*] *Corresponding Author, email address: peregoa@aston.ac.uk*

Nonlinear instabilities are responsible for spontaneous pattern formation in a vast number of natural and engineered systems ranging from biology to galaxies build-up. We propose a new instability mechanism leading to pattern formation in spatially extended nonlinear systems based on a periodic antiphase modulation of spectrally-dependent losses arranged in a zig-zag way: an effective filtering is imposed at symmetrically located wavenumbers $k$ and $–k$ in alternating order. The properties of the dissipative parametric instability differ from the features of the both key classical concepts of modulation instabilities: the Benjamin-Feir, and the Faraday instability. We demonstrate how dissipative parametric instability can lead to the formation of stable patterns in one and two-dimensional systems. The proposed instability mechanism is generic and can naturally occur or can be implemented in various physical systems.

Formation of patterns in nonlinear physical and biological systems gives the conceptually important idea how simple objects can self-evolve to complex structures through instabilities. Spontaneous pattern formation in a variety of nonlinear spatially extended systems is initiated by modulation instabilities (MI): the homogeneous state becomes unstable with respect to growing spatial modulation modes in a given range of wavenumbers [1]. Possibly the best known class of MI is the Benjamin-Feir (BF) instability, originally introduced in fluid dynamics [2,3] and later identified in different areas of science, such as plasmas [4], nonlinear optics [5-7] and other fields (see, for example, the review [1]). The physical essence of BF instability is that some spatial modulation modes with symmetric wavenumbers $k$ and $–k$ can synchronize with the strong homogeneous mode with $k=0$ due to a nonlinear frequency shift in self-focusing (modulationally unstable) media, and thus can experience exponential growth.

Another fundamental MI – ubiquitous in physics – is the Faraday instability, historically known even before the BF instability. This instability results from the periodic modulation in time of an appropriate dispersive parameter of the system [8]. Faraday unstable modes oscillate at half the frequency of the parametric forcing. The Faraday instability can be understood as a synchronization of the growing modes at $k$ and $–k$ with the homogeneous mode through the periodic parametric driving. Specifically, when a parameter is time-modulated at frequency $2\omega_0$, the modes grow if their wavenumbers $k$ and $–k$ satisfy the nonlinear dispersion relation $\omega_0=\omega(k)$.

The Faraday instability was observed in a variety of systems: originally in vertically shaken fluids [8], later in periodically modulated chemical systems [9], in vertically shaken granular media [10], in periodically modulated Bose-condensates [11,12] and in nonlinear fiber optics. In the latter case, the modulation of nonlinearity or dispersion in time (piecewise or in a continuous manner) can initiate instability [13-16] and lead to pattern formation [17-19]. Typically, the Faraday instabilities and patterns are studied in BF-stable systems. However, they can also appear as additional instabilities in BF-unstable cases [20].

In this Letter, we propose a new type of instability that we call dissipative parametric instability. While it shares some features with BF and the Faraday instabilities, the dissipative parametric instability is also very distinct from these two classical cases.

In various applications, both BF and Faraday instabilities and the associated nonlinear pattern formation can be described using a very generic model, the complex Ginzburg-Landau equation [21] (CGLE), which (in the case of one spatial dimension) reads:

$$\frac{\partial A}{\partial t} = \mu A + (b - id)\frac{\partial^2 A}{\partial x^2} + (ic - s)A|A|^2 \quad (1)$$

where $A(t,x)$ is the complex field amplitude distributed in space $x$ and evolving in time $t$, $\mu$ is the gain coefficient, $s$ and $c$ are the saturation and nonlinearity coefficients and $b$ and $d$ are diffusion and diffraction coefficients. In the case of Faraday instability, diffraction $d(t)$ and/or nonlinearity $c(t)$ are periodic functions of time. Note that the modulation in time of dissipative parameters, such as $\mu$, $s$ or the diffusion $b$ (which effectively acts as dissipation for large $k$ components), does not result in Faraday instability. In conservative systems, such as nonlinear fibers and Bose-Einstein condensates, both BF and Faraday instabilities are studied within the framework of the nonlinear Schrödinger equation (NLSE), the conservative limits of CGLE.

The linear stability of the homogenous solution of the CGLE with periodic coefficients can be studied using the Floquet stability analysis. The homogeneous state $A_{hs}=A_0 \exp(ic|A_0|^2 t)$, in which $\mu=s|A_0|^2$, is weakly perturbed by modulation modes $a_{+k}(t)\exp(ikx)$ and $a_{-k}(t)\exp(-ikx)$, such that $a_{+k}(t)$, $a_{-k}(t) \ll |A_0|^2$. Calculating numerically the amplitudes of perturbations after one modulation period, building a matrix-map of a resonator roundtrip, and diagonalizing it (see Supplemental Material [22] for details) allows to calculate the Floquet multipliers $F$. A mode $k$ is considered unstable when at least one of the absolute values of its multipliers is greater than 1. In order to visualize the instability spectrum, we plotted $Max(|F(k)|)$.

The BF instability, in the CGLE and in its conservative limit (NLSE), is a long-wave instability, because the band of unstable wavenumbers always extends from $k=0$; see Fig. 1(a). For particular systems, described e.g. by Manakov equations [23] or the Lugiato-Lefever equation [24], BF is not purely a long-wave instability but its spectrum can slightly detach from $k=0$. The Faraday instability is a short wave instability: the area of unstable modes is clearly detached from the axis $k=0$; see Fig. 1(c). There are multiple Faraday instability tongues. In the first tongue, the growing modes oscillate with half the frequency of the parametric drive, in the second tongue - with the frequency of the drive, and so on. Another fundamental difference is that BF unstable modes grow monotonically, as shown in Fig. 1(b), whereas the growth of Faraday unstable modes is oscillatory and synchronized with the parametric drive, as in Fig. 1(d) (see also Supplemental Material [22]).

The new type of instability – dissipative parametric instability – occurs in systems in which dissipative terms are periodically modulated in time in an antiphase (zig-zag) manner with respect to $k$ and $-k$ modes, Fig. 2(a).

First, the complex field evolves nonlinearly and homogeneously in time according to the CGLE with non-modulated coefficients. Next, spectral losses are imposed over the wavenumber range $-\Delta k$ at time instant $t=T_f/2$. Then a new stage of homogeneous nonlinear evolution takes place, followed by spectral losses over the wavenumber range $+\Delta k$ at $t=T_f$. Note that the unmodulated dissipation in the $k$-domain (constant diffusion coefficient $b$ in Eq. (1)) or the symmetrically (for $k$ and $-k$ modes) modulated dissipation does not result in any instability. Additionally, we would like to point out that the dissipative term remains positive on average at every instant of time, i.e. the losses are never converted into gain.

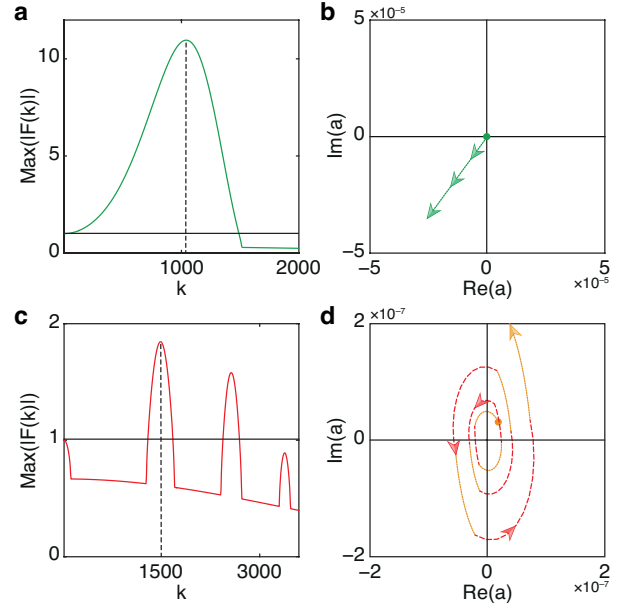

FIG. 1. (a, c) Floquet spectrum calculated using the Floquet stability analysis of the homogeneous solution of the CGLE, instability occurs above the horizontal continuous line. (b, d) The dynamics of complex amplitude $a(k)$ of the most unstable mode (indicated by dashed vertical lines on the instability spectrum) is calculated by direct integration of the CGLE. Arrows indicate the direction of temporal evolution. Parameters are $\mu=1$, $s=0.3$, $b=0.1\cdot 10^{-6}$, full integration time $T=1$. In the case of BF instability (a, b): $c=1$, $d=-3\cdot 10^{-6}$. In the case of Faraday instability (c, d): $c=4.85$, $d_1=5\cdot 10^{-6}$, $d_2=1\cdot 10^{-6}$, and a piecewise modulated diffraction coefficient is considered: $d=d_1$ for $0<t<0.2$ (orange line on (d)), then $d=d_2$ for $0.2<t<0.4$ (red line), and so on.

Such antiphase spectrally modulated losses can occur in periodic (cyclic) systems with spectrally shifted dissipative components. In the one-dimensional case, dissipative parametric instability can arise in transmission fiber systems, lasers or amplifiers, in which dissipative elements (such as filters) are imposed in alternating (zig-zag) order in frequency domain [25, 26]. A laser is a natural example of a system exhibiting dissipative parametric instability if the frequency reflectivity profile of one mirror is shifted with respect to that of the other mirror; see Fig. 2(b). Such detuning results in periodic antiphase losses at every half cavity round-trip.

Another possibility is to implement alternating losses in wavenumber domain. This could be realized in transverse nonlinear optics, such as self-imaging resonators [27] or self-imaging arrays of lenses, if the access to the far field distribution at different positions along the resonator is possible. Selective losses for the $+k$ and $-k$ components can be imposed by placing corresponding spatial filters, see Fig. 2(c). The dissipative parametric instability could also be implemented in dissipative Bose-Einstein atomic or exciton-polariton condensates in semiconductor microcavities [28]. In the first case velocity (momentum) resolved losses have to be imposed; in the second case Bragg mirrors with suitable reflectivity profiles must be used (Fig. 4(a), Supplemental Figs. 5(a) and 5(c)).

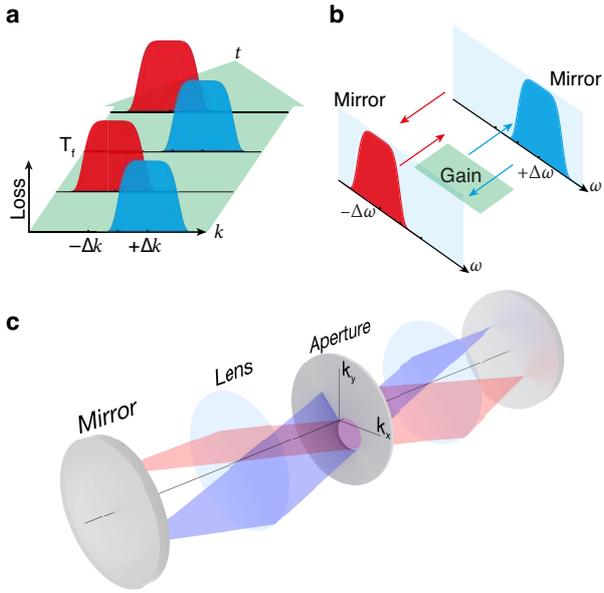

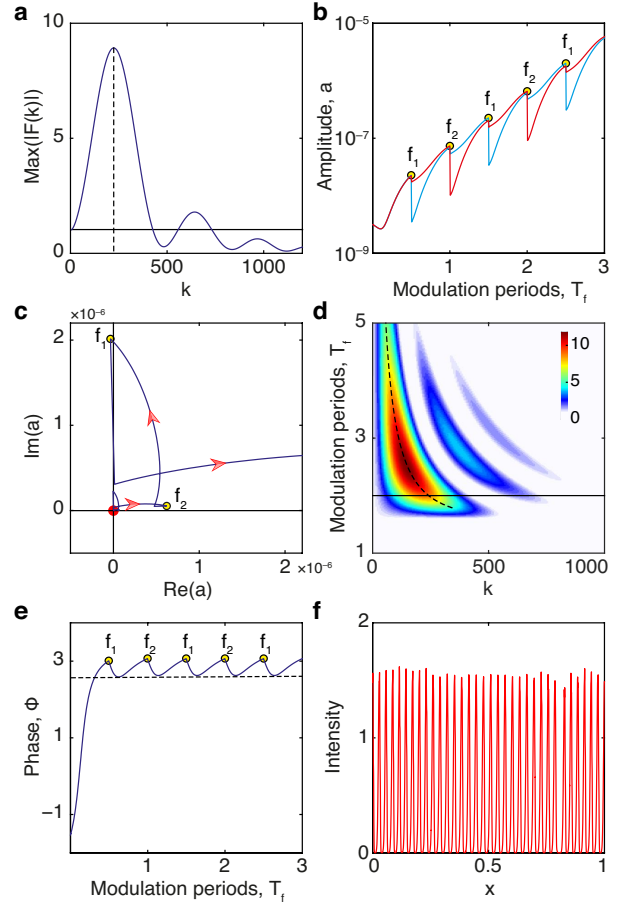

FIG. 2. (a) Dissipative parametric instability arises if periodic in time losses are introduced asymmetrically in the $k$-domain, so that the modes with only positive or negative wave vectors are damped every half of the period $T_f$. (b) Dissipative parametric instability can be realized by alternating losses in frequency domain, i.e. in a laser with detuned (in frequency) cavity mirrors. (c) A self-imaging resonator or self-imaging array of lenses with spatial filter displaced relative to the system's axis is another possibility.

We calculated the properties of the dissipative parametric instability in a system described by Eq. (1), modelling without loss of generality the dissipative elements as super-Gaussian spectral filters: $f_{1,2}(k)=\exp(-(k\pm k_0)^8/\sigma^8)$. The specific spectral shape of the filter function is not critical for the properties of the dissipative parametric instability. We performed the Floquet analysis for periodic in time and antiphase in $k$-domain losses (see Fig. 3). The parameters used in the calculations are $\mu=1$, $s=0.2$, $c=3.5$, $b=0.1\cdot 10^{-6}$, $d=5\cdot 10^{-6}$, $k_0=1822.1$ and $\sigma=1885$, modulation period is fixed, $T_f=2$, except in Fig. 3(d) where $T_f$ has been varied.

The dissipative parametric instability band starts from $k=0$, Fig. 3(a), which makes its spectrum similar to the BF-instability (compare with Fig. 1(a)). We stress that the system considered on average and in every instant of time remains in the BF-stable regime. At the same time, the dissipative parametric instability spectrum has several tongues, as in Fig. 3(a) and 3(d), which is characteristic to Faraday instability. We note that the dissipative parametric instability spectrum could be also tailored to make modes with small wavenumbers stable, or to modify the number of instability tongues by changing the dissipation function, such as its shift over frequency and the modulation period over time.

As in Fig. 3(b), on average, the amplitudes of the unstable modes grow exponentially, but oscillate synchronously with external forces like in the Faraday instability case, and unlike the monotonous evolution of BF instability (see Supplemental Fig. 1 [22] for comparison). The complex amplitudes of the modulation modes perform looping in the phase space synchronized with the external modulation of dissipation, as in Fig. 3(c). The evolution in the phase space for the dissipative parametric instability is different from the cases of both BF and Faraday instabilities.

FIG. 3. (a) Spectrum of dissipative parametric instability. The dashed line indicates the most unstable mode. (b) Evolution of the absolute values of the amplitudes of the most unstable modes $a(k)$ and $a(-k)$, red and blue lines, respectively. The losses are introduced at points $f_1$, $f_2$ etc. in time. (c) The complex amplitude of the mode $a(k)$ evolves in loops in phase space synchronized with external forcing. (d) Spectrum of dissipative parametric instability as a function of the modulation period $T_f$, the dashed line is the analytically estimated scaling law of the instability. (e) The generalized phase $\Phi$ locks to the optimum value (dashed line) at which the mode's amplitudes are growing at the fastest rate through periodic reset of the phase at instances of time at which the losses are applied. (f) Asymptotically stable pattern in one-dimensional system.

Despite the fact that modes with wavenumbers close to zero are unstable (see Figs. 3(a) and 3(d)), similarly to BF case, the dissipative parametric instability exhibits different scaling laws compared to BF instability. Indeed, whereas in the BF-instability case the instability spectrum does not scale over the modulation period (system length) [7], the scaling is well pronounced in the case of dissipative parametric instability, as depicted in Fig. 3(d). To characterize the scaling law, we phenomenologically assumed the parametric resonance condition, as for Faraday instability, by imposing that the first unstable mode oscillates in time at frequency $\omega_f/2$, with $\omega_f$ as frequency of the forcing and has wavenumber $k_{inst}$ related to $\omega_f/2$ through the dispersion relation. The resulting analytically derived scaling law (See Supplemental Material [22]) coincides well with the numerical calculation, Fig. 3(d).

How and why the dissipative parametric instability emerges becomes clear through the calculation of the generalized phase $\Phi=\phi_{+k}+\phi_{-k}-2\phi_0$, where $\phi_{+k}$ and $\phi_{-k}$ are the phases of the modes with wavenumbers $+k$ and $-k$, respectively, and $\phi_0$ is the phase of the homogeneous mode. For a BF stable system, if the dissipation is not modulated, the generalized phase $\Phi$ evolves freely over time, resulting in periodical growth and decay of amplitudes of modulation modes depending on the instantaneous phase in the way that the period-average amplitude remains the same. In the conservative limit, such periodically oscillating modes are well known under the name of Bogoliubov-De Gennes excitations - sound waves of a condensate [29].

The situation is completely different when the modes at $+k$ or $-k$ are periodically damped in zig-zag fashion. In this case, at the instant of time during which the damping is applied the generalized phase is reset to the value at which the amplitude is growing; see Fig. 3(e). Despite the increased dissipation on average, the exponential growth of the unstable modes sets in. We directly checked that such dynamics cannot be sustained if both modes were damped in phase. In this way, dissipative parametric instability is fundamentally different from the Faraday instability, where the modes at $+k$ and $-k$ are modulated in phase.

The dissipative parametric instability eventually leads to pattern formation. For one-dimensional systems, we provide an example of a stable pattern evolved from the homogenous solution, in Fig. 3(f). The character of final patterns crucially depends on nonlinearity through the saturation of the amplitudes of unstable modes. Typically, stable and regular periodic patterns are excited, however, depending on parameters, dynamic irregular patterns are observed, characterized by a permanent creation and annihilation of the pulse-like localized structures during the temporal evolution (see Supplemental Fig. 4 [22]).

The increase of the nonlinearity leads to decrease of the wavenumber of the modulation pattern; the increase of the modulation period $T_f$ and of dispersion coefficient $d$ have the same effect, in agreement with Supplemental Eq. 9 [22].

Furthermore an increase of the filters width or a reduction of their separation leads to a lower modulation wavenumber, however a minimum separation is needed in order to excite the instability.

More comments on the patterns characterization, stability and temporal evolution can be found in Supplemental Material [22].

The dissipative parametric instability is of generic nature and could also be realized in higher dimensional systems. An example is a two-dimensional system that is stable with respect to BF instability, and where we apply the profile of the dissipation function in a zig-zagging manner, as depicted in Fig. 4(a).

As a result, see Fig. 4(b), the dissipative parametric instability appears with a corresponding instability spectrum and leads to pattern formation. Different patterns could be obtained depending on the system parameters that vary from completely stable and regular modulation patterns, as in Fig. 4(c), to irregular ones, as in Fig. 4(d). The resulting periodic patterns in saturated regimes (when they are stable) are of wavenumbers within the area in $k$-space, where the losses are modulated. The dissipative parametric instability in 2D spatial systems could be controlled by managing the shape of the dissipation function with significant flexibility (further examples of two-dimensional patterns are reported in Supplemental Material [22]).

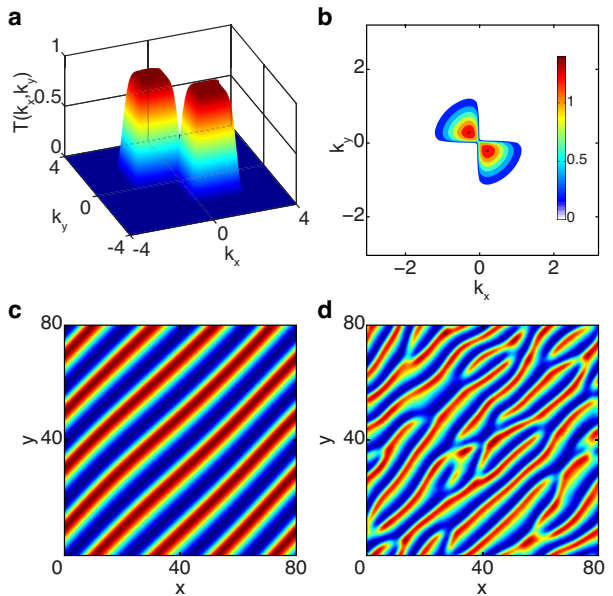

FIG. 4. (a) Zig-zagging losses in wavenumber space $(k_x,k_y)$, (b) the instability area in $(k_x,k_y)$ space as obtained by the Floquet analysis and (c) 2D intensity patterns. Parameters are $\mu=0.2$, $d=0.05$, $b=0.08$, $c=0.35$, $s=0.3$, $T_f=5\pi$, $\sigma=1.0905$. Losses are centred at $k_{0x}=-1$, $k_{0y}=+1$; (d) By setting $b=0$, the pattern becomes irregular in space and nonstationary in time.

In conclusion, we proposed and examined the dissipative parametric instability, a novel type of instability that can lead to pattern formation. The dissipative parametric instability occurs as a result of the periodic, in time, antiphase (zig-zagging) modulation of the spectral losses in the wavenumber (or frequency) domain. We have shown that this novel instability can lead to the formation of stable patterns in one and two-dimensional systems. The dissipative parametric instability is generic and can occur in various physical systems, including fiber optics, lasers and Bose-Einstein condensates.

We acknowledge support by the Spanish Ministerio de Educación y Ciencia, European FEDER project FIS2011-29731-C02-01, the ERC project ULTRALASER, the Russian Ministry of Education and Science (14.B25.31.0003), the Russian Foundation for Basic Research (15-02-07925), Presidential Grant for Young researcher (14.120.14.228-MK) and Dinasty Foundation. N.T. is supported by the Russian Science Foundation (14-21-00110). A. M. P. acknowledges support from the ICONE Project through the Marie Curie grant No. 608099.

# Supplemental Material: Pattern generation by dissipative parametric instability

A. M. Perego[1,2], N. Tarasov[1,3], D. V. Churkin[1,4,5], S. K. Turitsyn[1,4] and K. Staliunas[2,6]

[1] *Aston Institute of Photonic Technologies, Aston University, Birmingham B4 7ET, UK*
[2] *Departament de Fisica i Enginyeria Nuclear, Universitat Politècnica de Catalunya, E-08222, Barcelona, Spain*
[3] *Institute of Computational Technologies SB RAS, Novosibirsk, 630090, Russia*
[4] *Novosibirsk State University, Novosibirsk, 630090, Russia*
[5] *Institute of Automation and Electrometry SB RAS, Novosibirsk, 630090, Russia*
[6] *Institució Catalana de Recerca i Estudis Avançats, Passeig Lluis Companys 23, E-08010, Barcelona, Spain*

In order to perform the numerical stability analysis of Eq. (1), we first calculated the homogeneous solution. Then we added a small complex perturbation to each spectral mode of the spectrum and integrated the CGLE for one modulation period $T_f$. To be more specific, a 4 by 4 transfer matrix $M$ was obtained for each mode pair $+k$ and $-k$, whose first and second row entries are the real and imaginary parts of the modes $+k$ and $-k$ amplitudes after the evolution of real and imaginary perturbations to mode $k$. The third and fourth rows of $M$ contain the real and imaginary parts of $+k$ and $-k$ mode amplitudes, respectively, after the evolution of real and imaginary perturbations of mode $-k$. The resulting modes' amplitudes were normalized to the initial perturbation's absolute value.

The diagonalization of matrix $M$ provides a set of four eigenvalues $F(k)$ for modes $+k$ and $-k$: the so-called Floquet multipliers. A mode $k$ is considered unstable when at least one of the absolute values of its eigenvalues is greater than 1. To visualize the instability spectrum, we plotted $Max(|F(k)|)$ – the maximal absolute value of the Floquet multipliers – for each mode. As the instability spectrum is symmetric, only the positive part the spectrum ($k>0$) has been plotted.

The mechanism of dissipative parametric modulation instability differs from those of the classical Benjamin-Feir and Faraday instability due to its *antiphase* modulation dynamics depicted in Fig. 3(b) of the main article. Specifically, the amplitudes of modulation modes symmetrically located at $+k$ and $-k$, respectively, both grow on average over time. During this increase, however, their amplitudes are not equal at every instant point of evolution due to the action of the spectrally dependent losses. This feature clearly distinguishes the reported dissipative parametric instability from the BF and the Faraday ones.

In the case of Faraday instability, the growth process is synchronized with the external forcing. In the Benjamin-Feir case, since no periodic forcing is applied, the growth is due to the increase of the small perturbations during the evolution (see Supplemental Fig. 1). Hence, the synchronization with the external forcing is a common feature of both Faraday and dissipative parametric instability.

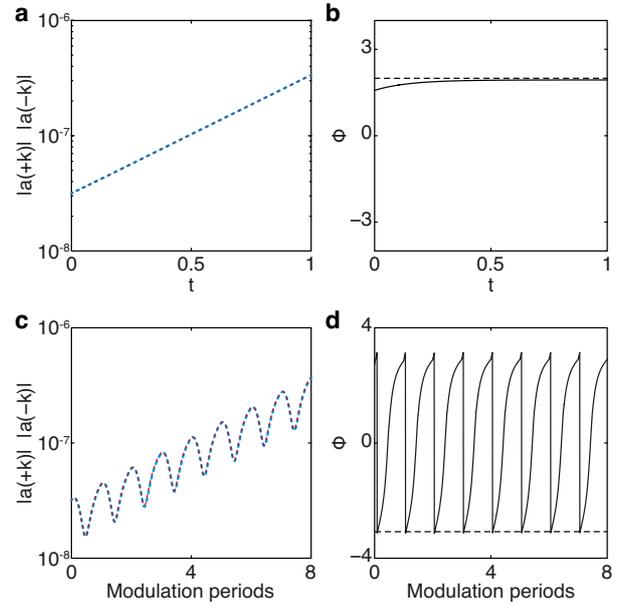

Supplemental FIG. 1. Growth process of the symmetrically (in wavenumber space) located maximally unstable modes $a(+k)$ and $a(-k)$ (blue and dashed red line) (a) and generalized phase (b) for the BF instability. The same for the Faraday instability: modes (c) and generalized phase (d). Dashed lines in (b) and (d) correspond to the optimum value of the generalized phase for synchronization with the homogeneous mode. For Faraday instability, the oscillatory growth process is synchronized with the external forcing. The parameters used are those considered in Fig. 1 of the main article.

We provide now a heuristic explanation of the growth of the unstable modes in the dissipative parametric instability of the CGLE, showing how, in presence of the alternating (zig-zag) damping, the coupling between modes can provide the energy necessary for the growth. Let us consider the CGLE (Eq. (1) of the main article). After perturbation of the homogeneous solution choosing the following *ansatz* for the field $A(t,x)=A_0 exp(ic|A_0|^2 t)[1+a_+ exp(ikx)+a_- exp(-ikx)]$; and linearization of CGLE with respect to the small perturbations, the evolution equation for the modulation mode $a_{+k}(t)$, reads:

$$\frac{\partial a_+}{\partial t} = \mu a_+ - bk^2 a_+ + idk^2 a_+ + ic(a_+ + a_-^*)A_0^2 - s(2a_+ + a_-^*)A_0^2. \quad (SE\ 1)$$

The (+) mode is coupled to the mode $a_{-k}^*(t)$:

$$\frac{\partial a_-^*}{\partial t} = \mu a_-^* - bk^2 a_-^* - idk^2 a_-^* - ic(a_+ + a_-^*)A_0^2 - s(2a_-^* + a_+)A_0^2. \quad (SE\ 2)$$

The solution of Supplemental Eqs. (1) and (2) can be sought in the form of exponentially decaying oscillations (Bogoliubov-De Gennes excitations): $a(t)=\exp(Dt)\,[a_1\cos(\omega_B t)+a_2\sin(\omega_B t)]$. The frequency of the oscillations $\omega_B$ and the damping rate $D$, are given by the imaginary and real part of the eigenvalue spectrum of the CGLE, respectively:

$$\lambda_\pm = -\mu - bk^2 \pm \sqrt{-d^2 k^4 - 2cdk^2 \mu/s + \mu^2}. \quad (SE\ 3)$$

In the limit $d^2k^4+2cdk^2\mu/s > \mu^2$, the frequency and the damping coefficient are, respectively:

$$\omega_B = \sqrt{d^2 k^4 + 2cdk^2 \mu/s - \mu^2} \quad (SE\ 4)$$

$$D = -\mu - bk^2 \quad (SE\ 5)$$

and the corresponding solutions of Supplemental Eqs. (1) and (2) read:

$$a_+(t) = \exp[(-\mu - bk^2)t]\{\tilde{a}_+ \cos(\omega_B t) + [\tilde{a}_-^*/\omega_B(ic\mu/s - \mu) + \tilde{a}_+/\omega_B(idk^2 + ic\mu/s)]\sin(\omega_B t)\} \quad (SE\ 6)$$

$$\tilde{a}_-^*(t) = \exp[(-\mu - bk^2)t]\{\tilde{a}_-^* \cos(\omega_B t) + [\tilde{a}_+/\omega_B(-ic\mu/s - \mu) + \tilde{a}_-^*/\omega_B(-idk^2 - ic\mu/s)]\sin(\omega_B t)\} \quad (SE\ 7)$$

with $a_+(0) = \tilde{a}_+$ and $a_-^*(0) = \tilde{a}_-^*$. We can obtain the temporal evolution for the mode $a_{-k}(t)$ by taking the complex conjugate of Supplemental Eq. (7):

$$a_-(t) = \exp[(-\mu - bk^2)t]\{\tilde{a}_- \cos(\omega_B t) + [\tilde{a}_+^*/\omega_B(ic\mu/s - \mu) + \tilde{a}_-/\omega_B(idk^2 + ic\mu/s)]\sin(\omega_B t)\} \quad (SE\ 8)$$

where $\tilde{a}_- = a_-(0)$.

The amplitudes of the excitations exponentially decay asymptotically, oscillating at frequency $\omega_B$ as illustrated in Supplemental Fig. 2(a). However, when the initial amplitude of one mode, say $a_-$, is much lower than the amplitude of the other one, $a_+$, then the amplitude of $a_-$ grows due to the coupling, as depicted in Supplemental Fig. 2(b).

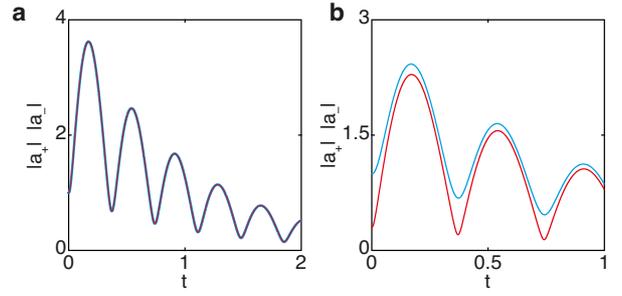

Supplemental FIG. 2. Temporal evolution of Bogoliubov-De Gennes modes for $k = \pm 100 \times 2\pi$ obtained evaluating Supplemental Eqs. (6) and (8) with the same parameters as in Fig. 3 of the main article. The excitations experience oscillatory behavior with asymptotic decay of the amplitudes (blue and over-imposed dashed red line) (a) when the initial conditions are equal; in particular, we have chosen $\tilde{a}_+ = \tilde{a}_- = 1$. When one mode is damped, a rapid growth of its amplitude occurs, as shown in (b); in this case $\tilde{a}_+ = 1$ and $\tilde{a}_- = 0.3$. Alternating the damping of modes $\pm k$ with a temporal periodicity, which allows the successive growth of the damped mode, leads to the average growth of both sidebands, resulting in the dissipative parametric instability.

When the losses for modes $a_-$ and $a_+$ are introduced in an alternating way and with a period large enough to allow for the growth of the damped mode – but not too large – to avoid the asymptotic decay, then an average growth of $a_-$ and $a_+$ occurs.

The evolution described in terms of Supplemental Eqs. (6) and (8) is valid in the linear regime, when the quadratic terms in the mode amplitudes are negligible. In order to describe the nonlinear dynamics, numerical integration of the master Eq. (1) of the main paper is required. Nevertheless, the linear analysis presented above sheds light on how the instability develops before entering the nonlinear regime, where the sidebands amplitudes are no longer small and the saturation process takes place. In principle, the instability can develop as a result of periodically imposed losses only on one mode (say with wavenumber $k$) or on a spectral region (say $+\Delta k$), this kind of excitation could not lead to pattern formation – only to a frequency shift in the spectrum. In order to achieve pattern formation, the spectral zig-zag modulation configuration is required.

Motivated by the synchronization between the growing modes and the external forcing shown in Fig. 3(b) of the main article, we present here an analytical estimate of the wavenumber of the maximally growing mode of the dissipative parametric instability (dashed black line in Fig. 3(d)). This is done by imposing the parametric resonance condition to the dispersion relation of the dissipative Bogoliubov modes of the CGLE. This condition assumes that the first excited mode has a wavenumber that is related, via the dispersion relation $\omega(k)$, to a temporal frequency equal to half of the forcing one. Starting from the instability spectrum of the CGLE, the dispersion relation is given by Supplemental Eq. (4). In the long wave limit, $2cdk^2\mu/s \gg d^2k^4$, Supplemental Eq. (4) simplifies to:

$\omega_B = \sqrt{2cdk^2 \mu/s - \mu^2}$, which allows straightforwardly to estimate, for $\mu \leq \pi/T_f$, the wavenumber of the first excited mode $k_{inst}$ by imposing the parametric resonance condition:

$$k_{inst} \approx \frac{(\pi/T_f)}{\sqrt{2cd\,\mu/s}}. \qquad \text{(SE 9)}$$

Supplemental Eq. (9) is generic and gives the estimate of the first unstable mode for the parametric instabilities. In the presence of strongly detuned filters, the homogeneous field intensity $|A_0|^2$ is not exactly equal to the nominal value $\mu/s$, since the strongly detuned filters can damp the homogeneous mode. However, we have checked that the minor damping of the homogeneous mode due to the detuned filters is not a necessary condition for the development of the dissipative parametric instability. In the absence of filters, selective and alternate damping of modes placed at $\pm k$ leads to their average growth. To plot the theoretical prediction (on Fig. 3(d)), we calculated $k_{inst}$ from Supplemental Eq. (9), using the intensity numerically averaged over one modulation period, instead of $\mu/s$. In Supplemental Fig. 3(c) this scaling is compared with the one that results from Supplemental Eq. (9) evaluated with the nominal value of $\mu/s$.

We have also considered an instability map similar to the one shown in Fig. 3(d) of the main article, but obtained for fixed modulation period and varying $\mu$. The instability map depicted in Supplemental Fig. 3(a) shows the unstable region as a function of the average intensity, which differs from the nominal value $\mu/s$ (Supplemental Fig. 3(b)) for the reasons mentioned above.

Another distinctive feature of the dissipative parametric instability is the scaling of the wavenumber of the most unstable mode with respect to the amplitude of the background wave. Our calculations at different wave amplitudes $A_0$ indicate that the maximally growing wavenumber decreases with field intensity (Supplemental Fig. 3), as can be expected for Faraday instability. This phenomenon contrasts with the well-known BF instability scaling in which the wavenumber of the maximally unstable mode always increases with the amplitude of the homogeneous field, in other words, with nonlinearity.

Despite its phenomenological origin Supplemental Eq. (9) provides a useful tool for a qualitative (or semi-quantitative) description of the dissipative parametric instability.

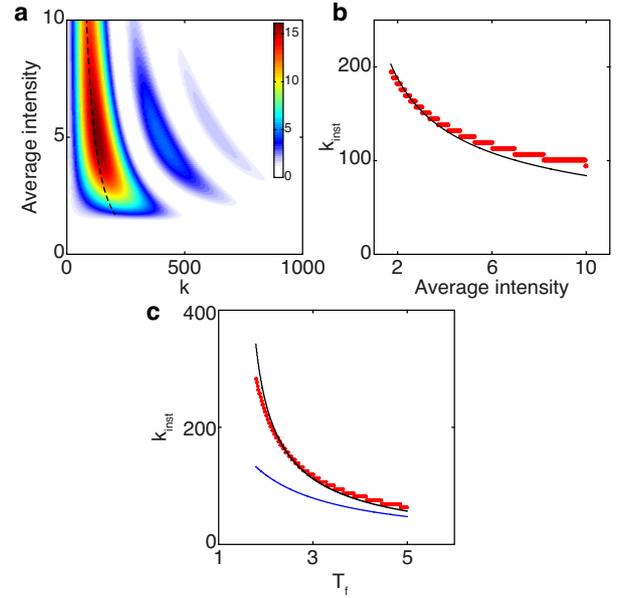

Supplemental FIG. 3. Instability map obtained varying $\mu$ from 0.5 to 2.142 and plotted in the wavenumber-average intensity space (a); the coloured regions correspond to instability; the remaining parameters are the same as in Fig. 3 of the main article. The scaling of the maximally unstable mode $k_{inst}$ versus the field intensity (b): red points are the results of Floquet analysis, the black line is Supplemental Eq. (9) with $\mu/s$ substituted by the effective average intensity calculated numerically. In (c), the scaling of the maximally unstable mode versus $T_f$, corresponds to Fig. 3(d) in the main article. The black line corresponds to Supplemental Eq. (9) using the average intensity, while the blue one is Supplemental Eq. (9) with the nominal value of the ratio $\mu/s$.

In the main article, we have provided an example of a pattern formation initiated by the dissipative parametric modulation instability. Even though a detailed study of the pattern stability conditions in the asymptotic nonlinear regime is beyond the scope of this study, we provide here two examples of regular and irregular patterns showing their temporal evolution. In Supplemental Fig. 4(a), a regular periodic pattern is depicted corresponding to the parameters used in Fig. 3 of the main article. Supplemental Fig. 4(b) shows the possibility of irregular patterns where repeated processes of creation and annihilation of spatial structures occur. The irregular pattern has been generated by reducing the detuning of the filters while retaining the remaining parameters as in the case of regular patterns. Supplemental Figs. 4(a) and 4(b) both depict a set of frames showing the spatial distribution of field intensity taken at the end of each modulation period, right after the second filter. We call a pattern "stable" if its shape remains unchanged for used long simulation time. Note, that this consideration does not prove true stability, but it gives a good indication of a possible stability of such patterns. As further check we have verified that performing the simulations which lead to pattern formation like the ones in Figs. 3(f) and 4(c), also in presence of additive noise, the resulting patterns remain unchanged. Stable patterns form when the noisy background which develops between the coherent structures is efficiently suppressed due to the combined action of nonlinear-dispersive spectral broadening and dissipative periodic filtering. If such suppression does not occur, then neighbor structures can grow between the already existing ones entering into competition with them with a related creation and annihilation process, as it is illustrated in Supplemental Fig. 4(b). The presence of diffusion helps the pattern stabilization.

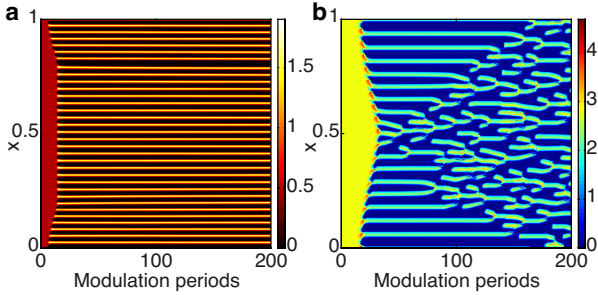

Supplemental FIG. 4. Temporal evolution of the one-dimensional patterns generated by the dissipative parametric instability. In (a), a stable pattern corresponding to the case of Fig. 3 of the main article is depicted; while in (b) the temporal dynamics of an unstable pattern shows continuous processes of creation and annihilation of coherent structures. Figure (b) has been obtained using $k_0 = 1570.8$, while keeping the remaining parameters as in (a).

Pattern formation through dissipative parametric instability in the two-dimensional system gives more freedom in the choice of the structure of the dissipative elements. Here we provide more details of the scheme illustrated in the main article and show them in Supplemental Figs. 5(a) and 5(b) where the dissipation function takes the form:

$$f_{1,2} = \exp[-(k_x \pm k_{0x})^2/\sigma^2]. \qquad (SE\ 10)$$

In contrast with the results presented in Fig. 4 of the main article, the patterns shown in Supplemental Fig. 5 are not tilted, because of the different shape of the dissipation in wavenumber space. A similar pattern, but with a spatial modulation along the orthogonal direction $y$, can be obtained by using the same dissipation function as in Supplemental Eq. (10), but replacing $k_x$ with $k_y$ and $k_{0x}$ with $k_{0y}$ (see Supplemental Figs. 5(c) and 5(d)).

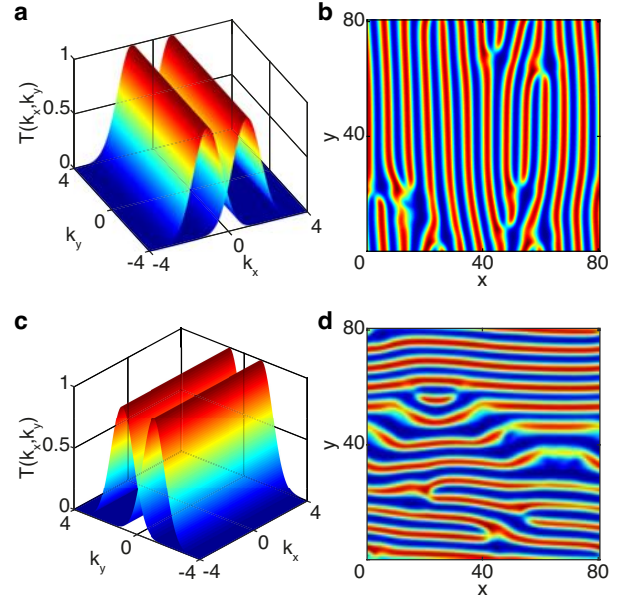

Supplemental FIG. 5. The two Gaussian transmission functions used to modulate the dissipation (a) and the corresponding 2-dimensional pattern created due to dissipative parametric instability (b) for $k_{0x}=1$. A $\pi/2$ rotation in $k$-space of the transmission function (exchange of $k_x$ with $k_y$ and $k_{0x}$ with $k_{0y}$) for $k_{0y} = 1$, leads to the generation of a pattern with a periodicity along the spatial direction $y$ (d). The parameters used are as follows: $\mu = 0.2$, $d = 0.05$, $b = 0.001$, $c = 0.35$, $s = 0.3$, $T_f = 5\pi$, and $\sigma = 1$.

We finally provide a phenomenological characterization of the patterns temporal evolution and functional shape. Once the pattern has appeared through progressive increase of the modulation of the homogeneous field background, the individual coherent structures which form the pattern evolve dynamically and periodically during each modulation period. The evolution in normal diffraction, in presence of gain, resembles the formation of similaritons in fiber amplifiers and leads to a considerable broadening associated with a modification of the original Gaussian shape into an almost parabolic one. The patterns shapes at the end of the nonlinear evolution just before the filter action are reported in Supplemental Fig. 6.

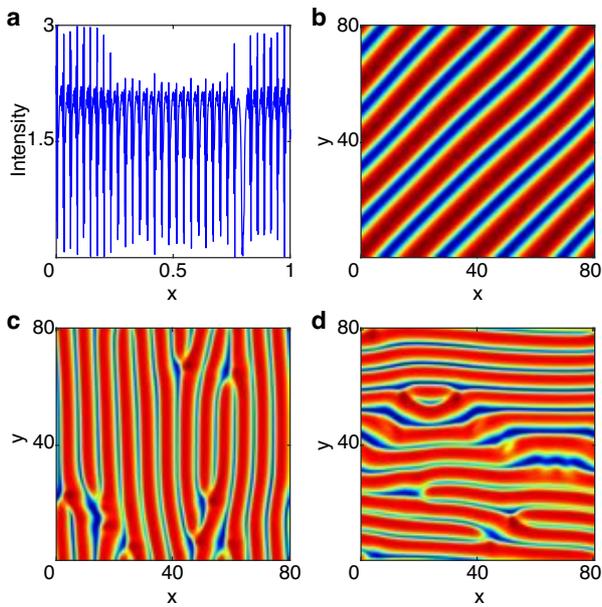

Supplemental FIG. 6. In (a) the 1-D pattern just before the filter action is depicted: the structures exhibit a clear broadening towards parabolic shape, while high frequency noise spikes are clearly visible between neighbour structures. In 2D we have the same broadening effect towards parabolic shape: (b), (c) and (d) are the corresponding intensity profiles before filter action for Fig. 4(c), Supplemental Figs. 5(b) and 5(d).

We have characterized the functional shape of individual structures fitting them with Gaussian and parabolic functions respectively after and before the action of the filter as it is clearly depicted in Supplemental Fig. 7.

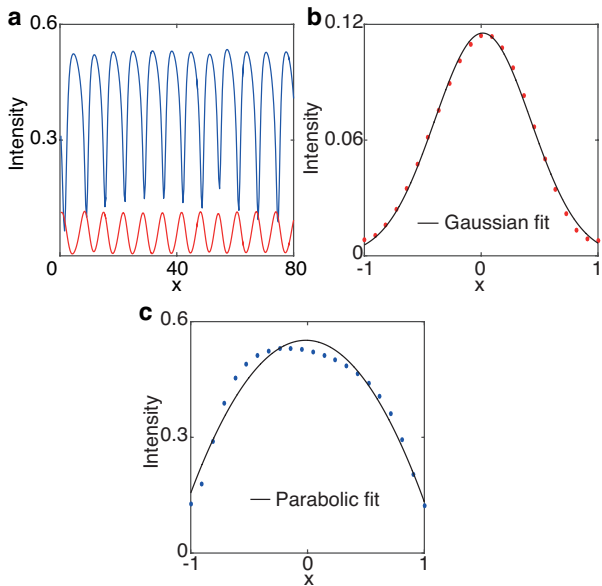

Supplemental FIG. 7. In (a) we present a section of the pattern depicted in Supplemental Fig. 5(b), in blue is the intensity profile before filtering while in red after filtering. In (b) and (c) are the fits of the single structures respectively after (Gaussian fit) and before (parabolic fit) the filter action.